# Mode conversion enhancement between silicon micro-slab and plasmonic nano-gap waveguides

Y. Liu[1,*] and Y. Lai[2]

[1]Department of Electrical and Computer Engineering, Texas A&M University, College Station, Texas, U.S.A.

[2]Department of Photonics, National Chiao-Tung University, Hsinchu, Taiwan.

[1,*]Corresponding author: ylgogogo@tamu.edu

**Foreword:** The experiment work of this project has been setup and done.

**Abstract:** We investigate a short (~1.5μm) partially-corrugated tapered waveguide structure for mode coupling from a silicon micro-slab to a plasmonic nano-gap waveguide at the optical communication frequency. More than 80% transmission efficiency is reported numerically for the first time. The result indicates that the corrugated waveguide structure should not only be helpful for realizing full on-chip silicon plasmonic devices but also a good choice for mode coupling enhancement from dielectric waveguides to plasmonic waveguides.

# 1. Introduction

With the advances of nano fabrication technologies, plasmonic waveguide devices have attracted intensive research interest in recent years mainly due to their strong optical confinement property at the scale that is much smaller than the free space optical wavelength. Such field confinement property provides a promising platform for the implementation of nano metalic devices for optical communication applications. Among all the required new technologies, efficient light coupling is one of the important issues for the design of gap plasmonic waveguides since the effective index of $TM_0$ plasmonic mode is higher than the material indices of the waveguiding structure, which makes waveguide mode couple to plasmonic mode difficult, especially when the difference of effective mode index between the conventional waveguide mode and the plasmonic mode is large. So far, there are a couple of literatures discussing the excitation of the plasmonic gap waveguide mode by using a plasmonic gap taper [1], a nano-antenna [2], and a multi-section coupler [3]. However, none of them are ideal for the mode coupling of metal-silicon-metal plasmonic gap waveguides due to the low coupling efficiency resulted by the aforementioned large effective mode index difference.

For better understanding, as shown in Fig. 1(a) and (b), the effective mode index $n_{eff}$ versus the waveguide width is plotted for the fundamental TM mode of the dielectric slab, the plasmonic $TM_0$ mode and the plasmonic $TM_2$ mode respectively. According to Fig. 1(b) and (d), one can observe that the fundamental TM mode of the silica slab is easier to couple into the plasmonic $TM_0$ mode than the plasmonic $TM_2$ mode as the effective index of the slab fundamental is closer to that of plasmonic $TM_0$ mode. On the contrary, from Fig. 1(a) and (d), one will expect that most power of the fundamental TM mode of the silicon slab will couple into the plasmonic $TM_2$ mode rather than the plasmonic $TM_0$ mode because its effective index is more close to that of the plasmonic $TM_2$ mode. This fact results in the low coupling efficiency for silicon based slab waveguide to plasmonic gap waveguide mode conversion with the methods [1-3] mentioned previously.

In microwave regime, corrugated metal structures have long been proposed and utilized as waveguide mode converters [4], surface wave assisted structures [5], slow-wave structures [6] or filters [7]. Recently, they draw attentions again in the emergent researches about slow-light and THz applications [8], for similar purposes such as dispersion controlling [9] and the so called "spoof" or "designer" surface plasmon assisted structures operating at low THz frequency [10-13], which are actually described in Electromagnetics textbooks [5-6] as surface waves existing on inductive corrugated surfaces formed by perfect electrical conductors (PEC).

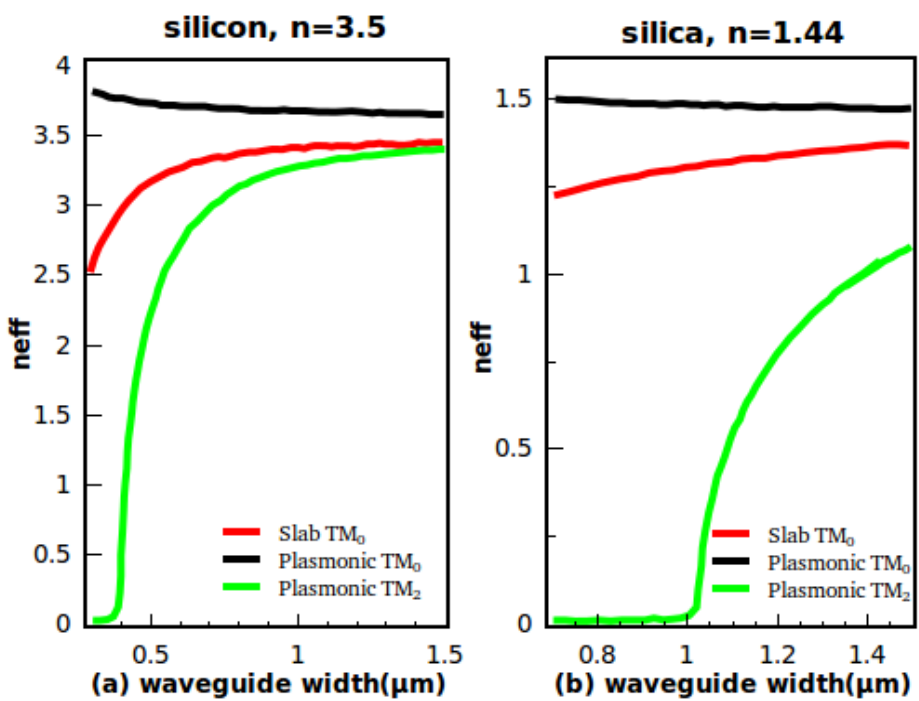

Fig. 1. Effective mode index of plasmonic TM0, plasmonic TM2 and slab TM0 mode for (a) silicon and (b) silica based slab waveguide and plasmonic gap waveguide. (c) The field distribution for the silicon based plasmonic gap waveguide coupling, one can observe that more power couple to plasmonic TM2 mode than plasmonic TM0 mode. (d) The field distribution for the silica (SiO2) based plasmonic gap waveguide coupling, one can observe that more power couple to plasmonic TM0 mode rather than plasmonic TM2 mode.

As mentioned above, a corrugated metallic waveguide can serve as a mode converter [4] due to its dispersion engineerable structure and low attenuation characteristics [14]. Unlike open corrugated structures [11, 12,13,15], the waveguide dispersion of a corrugate waveguide can extend across the light line [4, 6, 9, 14,16], which could be utilized in the coupler design to couple waveguide guided modes to evanescent modes, as shown in Fig. 2. Nevertheless, at the optical frequency, metal is no longer a perfect conductor and the signal propagation loss is considerable especially when the corrugated metal structure is incorporated in the design. Therefore, whether a metallic corrugated coupler with high coupling efficiency is feasible for guided mode to plasmonic mode conversion at the optical regime still requires further investigation. Recently, grooved metal sidewalls [17] (or another name, corrugated horn structure [18]) with 6.2 μm input opening for metal-silicon-metal plasmonic gap waveguide coupling is reported with maximum 72% coupling efficiency. However, the coupling is achieved by exciting SPPs (Surface Plasmon Polaritons) at metal surfaces so that the grove number will affect the transmission dramatically (Fig. 2 of [17]). In addition, the structure requires that (1) SPPs generated by adjacent grooves are in phase and (2) the incident lights falling on each groove are in phase for higher coupling efficiency; therefore the metal corrugated surfaces have to be tiled at a specific angle and the groove distance has to be kept at $2\lambda_{spp}$ once the metal material is decided. These restrictions may somehow constrain the coupler design freedom.

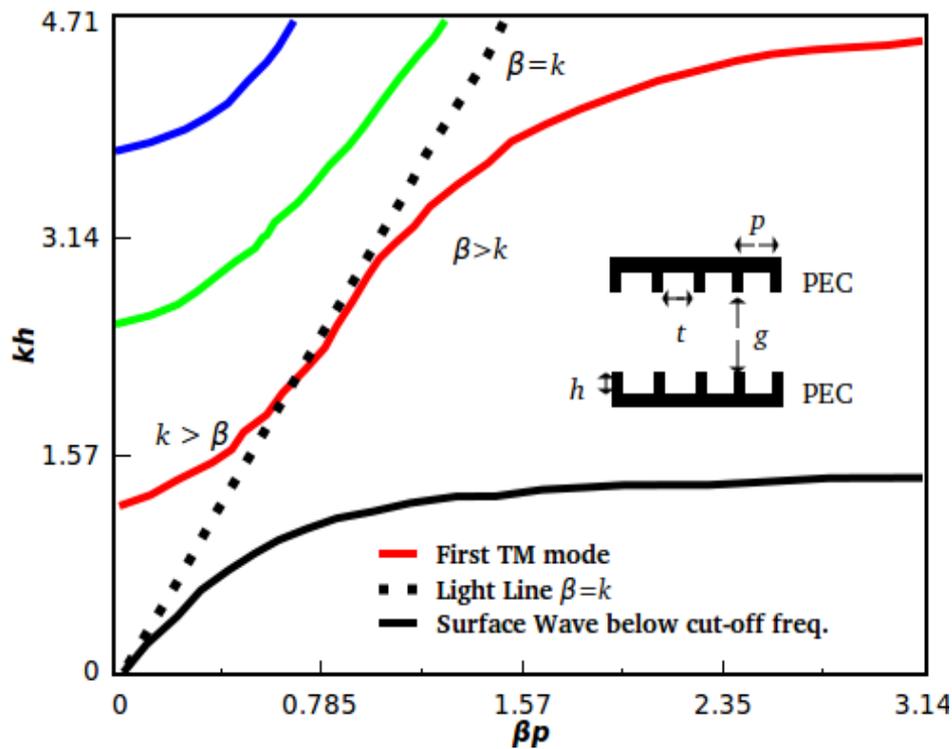

**Fig. 2. Dispersion diagram of a corrugated parallel-plate waveguide. As shown in the red line, the waveguide can have modes with both β>k and β <k.**

In this paper, a short (~1.5μm) partially gold corrugated tapered waveguide for mode coupling enhancement at the 1550nm optical communication wavelength between an all-silicon 1.25 μm micro-slab and a plasmonic nano-gap waveguide is designed and analyzed for the first time. The coupling efficiency is examine to be able to reach 86%~96% with plasmonic gap sizes at 20 nm, 50nm and 300nm respectively, which is comparable to or even higher than the previously referred non-silicon and silicon based plasmonic gap waveguide coupling devices. Finally, for comparison, silver is also adopted in the design for 50nm plasmonic gap waveguide coupling and the simulation result shows that around 92% coupling efficiency can be achieved by using the corrugated taper without the need to set the tilting angle of metal surfaces to be 32º, which is much higher than the coupling efficiency (~50%) of the aforementioned grooved metal sidewalls [17] and is a direct proof that the coupling mechanism of the corugated waveguide is different from that of [17].

## 2. Design and Analysis

Fig. 2 shows a PEC periodic corrugated metallic parallel plate waveguide and its dispersion diagrams [6]. The period of the corrugation, the depth of the tooth, the width of the dielectric tooth, and the vertical distance between metal teeth are represented by p, h, t and g respectively. One can see the corrugated waveguide can have modes across light line, as depicted in [4,6,14,16]. Although the dispersion curves in Fig. 2 is based on PEC corrugation as a periodic system (p~$\lambda$g) and its dispersion relation is different from PEC corrugation as a uniform systems (p<<$\lambda$g) [6], the relation between waveguide dispersion and the structure parameters (p, h, t and g) for these two cases still share the same features. Fig. 3 shows the partially-corrugated taper for the silicon filled plasmonic gap waveguide coupling. The width of the silicon slab $w_1$ is set to be 1.25 μm and the gap of the plasmonic metal-silicon-metal waveguide $w_2$ is set to be 20 nm in Fig. 3(a), 50 nm in Fig. 3(b) and 300 nm in Fig. 3(c) respectively. In the design simulation, the relative permittivity for silicon is $\varepsilon_{si}$ = 12.25 while the complex relative permittivity of the metal (gold) is $\varepsilon_{Au}$ = -93.0+11.0i determined by the Lorenz-Drude model [19]. For the concerns of numerical accuracy and stability, the simulation work is done by the Finite Element Method (FEM), which is more suitable for plasmonic related problems than the Finite-Difference-Time-Domain (FDTD) method because no material averaging is done at the material interface in FEM.

In the design, the corrugation period p (~0.14μm) is comparable to the effective wavelength λg (~0.44μm) of the slab fundamental mode, therefore the formulation for uniform system is not applicable here. In addition, since only a couple of corrugation periods are incorporated in the design, strictly speaking the corrugated taper cannot be considered as a periodic structure. Nevertheless, the dispersion relation for periodic systems still can provide a basic idea to decide the size of p, h and a for the effective mode index of the corrugated waveguide that matches the effective index of the fundamental slab $TM_0$ mode at the input end with g=1.25 μm and matches the effective index of the plasmonic $TM_0$ with g ~= 0.4 μm around which the plasmonic $TM_2$ mode is cut off as shown in Fig. 1. The reason why the taper is made partially corrugated is that only plamonic $TM_0$ mode exists when the metallic gap size drops below ~0.4 μm, as illustrated in Fig. 1(a); under this situation, the loss will increase when the plasmonic $TM_0$ mode keeps propagating on corrugated surfaces as fields of the mode concentrate on metal surfaces. In the design, p, t and the taper length are set to be around 0.14 μm, 0.1 μm and 1.5 μm respectively. With given β and g, h is then estimated by the following eq. (1) [6].

$$\frac{p}{t}\frac{\sqrt{\beta^2-k^2}}{k}\frac{tanh\left(\frac{g}{2}\sqrt{\beta^2-k^2}\right)}{sinc^2\left(\beta\frac{t}{2}\right)} = tan(kh) \qquad (1)$$

The value h for each tooth is fine tuned to get better performance. Based on the FEM simulation, the coupling efficiency for 20nm, 50nm and 300nm metal-silicon-metal gap can reach ~87%, ~89% and ~96% respectively, which proves that the corrugated waveguide can serves as the waveguide mode converter at optical frequency with low attenuation since most of the fields in the corrugated taper are still concentrated at the waveguide center with effective mode index higher than that of silicon, as shown in Fig. 3.

Based on the same procedure described previously, for performance comparison, silver is adopted in the design with relatvie permitivity $\varepsilon_{Ag}$ = -143.0+9.0i [19], of wich the imaginary part (counts for loss) is higher than that used in [17], i.e. $\varepsilon_{Ag}$ = $(0.144+11.366i)^2$ = -129.16522 + 3.273408i. The simulation result shows the coupling efficiency is ~92%, which is much higher than the case for 50nm gap coupling (~50 %) in [17]; the taper tilting angle of the structure shown in Fig. 4. is 21.8º and the groove distance is 0.15μm (< $2\lambda_{spp}$ ~= 1.48mm, according to data provided in [17]), which means the coupling mechanism of the corrugated structure reported in this paper is different from that of [17], although both strucutres look similar.

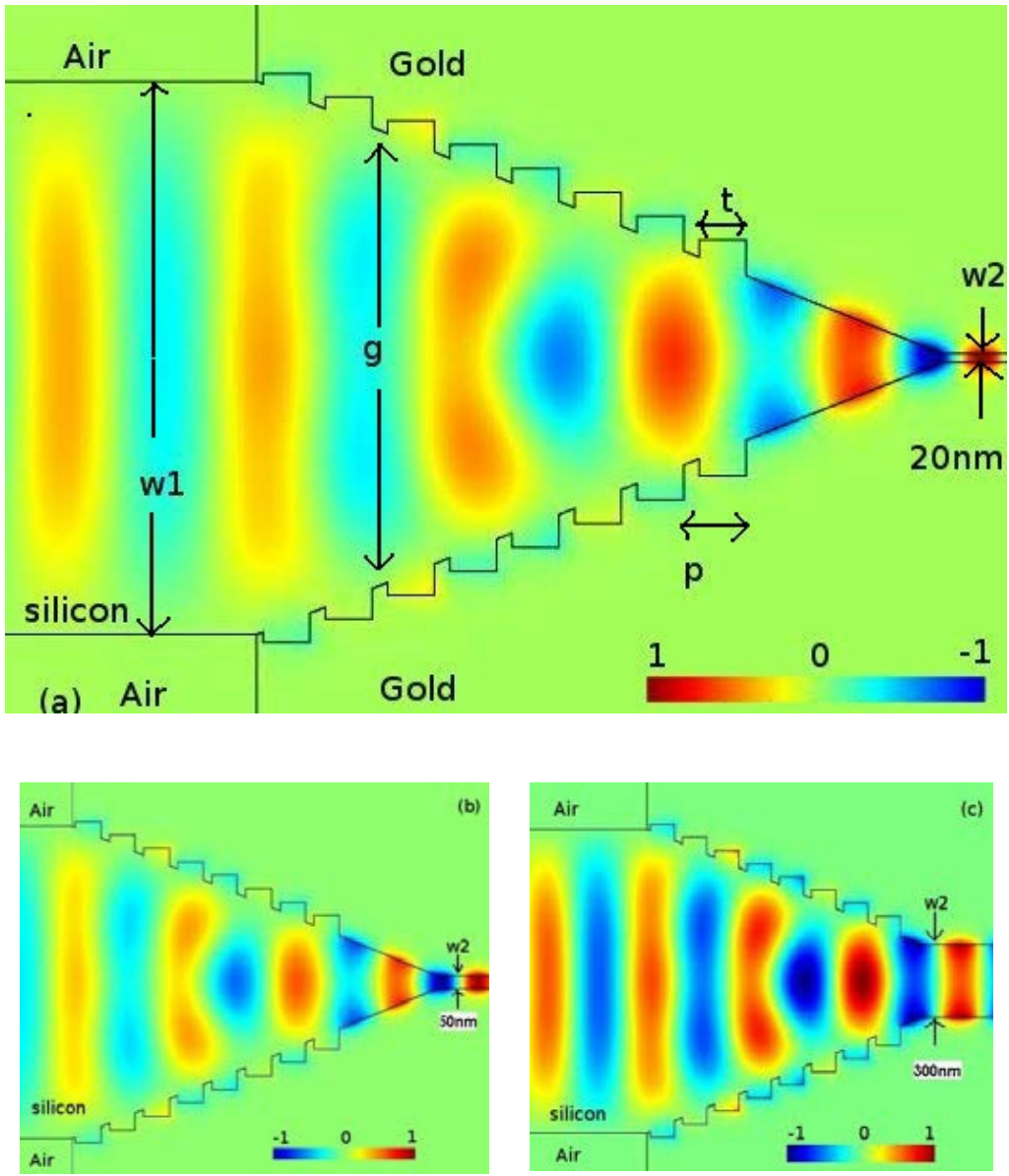

Fig. 3. 1.25μm (w1) silicon slab to plasmonic gap waveguide coupling for w2= (a) 20nm, (b) 50nm and (c) 300nm plasmonic waveguide width. Hz field is shown here.

## 3. Conclusion

To conclude, we have reported and designed a short (~1.5μm) coupler for efficient light coupling between a silicon micro-slab and a nano metal-silicon-metal plasmonic waveguide based on the dispersion engineering of the partially corrugated waveguide for dielectric slab guided mode to plasmonic guided mode matching. Numerical 86%~96% coupling efficiency for different gap sizes at the optical communication frequency is reported for the first time, which is comparable to or even higher than the previously referred performance of non-silicon and silicon based plasmonic gap waveguide coupling and should be helpful for realizing full on-chip silicon plasmonic devices. At the end, we point out that the coupling mechanism of the proposed structure in this paper is different from that of [17], which is achieved by exciting Spps on grooved metal surfaces.

## Acknowledgment

The author would like to acknowledge Mr. Thompson, Mr. Gray from bp and Mr. Fwu from SGI for providing workstations;  Texas A&M Supercomputing Facility and Prof. Liwen Shih of University Houston-Clear Lake for providing HPC.